\documentclass[sigconf,screen,9pt]{acmart}

\AtBeginDocument{%
  }

\copyrightyear{2026}
\acmYear{2026}
\setcopyright{cc}
\setcctype{by}
\acmConference[DAC '26]{63rd ACM/IEEE Design Automation Conference}{July 26--29, 2026}{Long Beach, CA, USA}
\acmBooktitle{63rd ACM/IEEE Design Automation Conference (DAC '26), July 26--29, 2026, Long Beach, CA, USA}
\acmDOI{10.1145/3770743.3806387}
\acmISBN{979-8-4007-2254-7/2026/07}

\usepackage{makecell}
\usepackage{threeparttable}
\usepackage{multirow}
\usepackage{array}
\usepackage{subcaption}
\usepackage{textcomp}
\usepackage{enumitem}
\usepackage{tcolorbox}
\usepackage{tabularx}

\newcommand{\papername}[1]{\texttt{AccelCIM}}

\newcounter{takeawaycnt}
\newcommand{\takeaway}[1]{%
    \stepcounter{takeawaycnt}
    \begin{tcolorbox}[
        colframe=black,
        colback=white,
        sharp corners,
        boxrule=0.4pt,
        boxsep=3pt, left=2pt, right=2pt, top=2pt, bottom=2pt,
        before skip=6pt plus 2pt minus 2pt,
        after skip=6pt plus 2pt minus 2pt
    ]
        \textbf{Takeaway \#\thetakeawaycnt:} #1
    \end{tcolorbox}%
}

\usepackage{tikz}
\newcommand{\blackcircle}[1]{%
  \tikz[baseline=(char.base)]{
    \node[shape=circle,draw,fill=black,text=white,inner sep=0.2pt,font=\footnotesize] (char) {#1};
  }%
}

\newcommand{\paperauthors}{%
  \shortstack[c]{%
    Chenhao Xue\textsuperscript{1,6,${\dagger}$}, 
    Yukun Wang\textsuperscript{2,6,${\dagger}$}, 
    An Guo\textsuperscript{3}, 
    Yuhui Shi\textsuperscript{3}, 
    Jinwei Zhou\textsuperscript{4},
    Xiping Dong\textsuperscript{1,6}, 
    Yihan Yin\textsuperscript{1,6}, \\
    Yuanpeng Zhang\textsuperscript{1,6}, 
    Tianyu Jia\textsuperscript{1,6}, 
    Wei Gao\textsuperscript{5},
    Qiang Wu\textsuperscript{5}, 
    Xin Si\textsuperscript{3}, 
    Jun Yang\textsuperscript{3}, 
    Guangyu Sun\textsuperscript{1,6,${\ddagger}$}%
  }%
}

\newcommand{\paperaffiliations}{%
  \textsuperscript{1} School of Integrated Circuits, Peking University, Beijing, China \par
  \textsuperscript{2} School of Electronics Engineering and Computer Science, Peking University, Beijing, China \par
  \textsuperscript{3} School of Integrated Circuits, Southeast University, Nanjing, China \par
  \textsuperscript{4} School of Integrated Circuits, Anhui Polytechnic University, Wuhu, China \par
  \textsuperscript{5} Houmo AI, Nanjing, China \par
  \textsuperscript{6} Beijing Advanced Innovation Center for Integrated Circuits, Beijing, China%
}

\makeatletter
\def\@mkauthors{%
  \begingroup
  \hsize=\textwidth
  \global\setbox\mktitle@bx=\vbox{%
    \noindent\unvbox\mktitle@bx\par\vskip2pt
    \centering
    {\@authorfont \paperauthors\par}
    \vskip4pt
    {\@affiliationfont \paperaffiliations\par}
    \vskip10pt
  }%
  \endgroup
}
\makeatother

\begin{document}
\title{AccelCIM: Systematic Dataflow Exploration for SRAM Compute-in-Memory Accelerator}
\orcid{0000-0001-9721-247X}
\author[C.~Xue]{Chenhao Xue}
\thanks{
This work is partially supported by the New Generation Artificial Intelligence-National Science and Technology Major Project (2025ZD0122105), Beijing Natural Science Foundation (L243001), 111 Project (B18001), NSFC under Grant 62522403, 62204036.}
\affiliation{%
  \institution{School of Integrated Circuits, Peking University}
  \city{Beijing}
  \country{China}
}

\author[Y.~Wang]{Yukun Wang}
\orcid{0009-0001-0769-1321}
\thanks{$^{\dagger}$ Equal contribution, email: \href{mailto:xch927027@pku.edu.cn}{xch927027@pku.edu.cn}, \href{mailto:2200012741@stu.pku.edu.cn}{2200012741@stu.pku.edu.cn}}
\affiliation{%
  \institution{School of Electronics Engineering and Computer Science, Peking University}
  \city{Beijing}
  \country{China}
}

\author[A.~Guo]{An Guo}
\orcid{0000-0003-4677-6114}
\affiliation{%
  \institution{School of Integrated Circuits, Southeast University}
  \city{Nanjing}
  \country{China}
}

\author[Y.~Shi]{Yuhui Shi}
\orcid{0009-0002-2933-1308}
\affiliation{%
  \institution{School of Integrated Circuits, Southeast University}
  \city{Nanjing}
  \country{China}
}

\author[J.~Zhou]{Jinwei Zhou}
\orcid{0009-0002-9637-4658}
\affiliation{%
  \institution{School of Integrated Circuits, Anhui Polytechnic University}
  \city{Nanjing}
  \country{China}
}

\author[X.~Dong]{Xiping Dong}
\orcid{0009-0001-8720-5804}
\affiliation{%
  \institution{School of Integrated Circuits, Peking University}
  \city{Beijing}
  \country{China}
}

\author[Y.~Yin]{Yihan Yin}
\orcid{0009-0002-3848-3426}
\affiliation{%
  \institution{School of Electronics Engineering and Computer Science, Peking University}
  \city{Beijing}
  \country{China}
}

\author[Y.~Zhang]{Yuanpeng Zhang}
\orcid{0009-0007-4253-4747}
\affiliation{%
  \institution{School of Integrated Circuits, Peking University}
  \city{Beijing}
  \country{China}
}

\author[T.~Jia]{Tianyu Jia}
\orcid{0000-0001-6645-509X}
\affiliation{%
  \institution{School of Integrated Circuits, Peking University}
  \city{Beijing}
  \country{China}
}

\author[W.~Gao]{Wei Gao}
\orcid{0009-0006-6881-5224}
\affiliation{%
  \institution{Houmo AI}
  \city{Nanjing}
  \country{China}
}

\author[Q.~Wu]{Qiang Wu}
\orcid{0009-0009-8981-2876}
\affiliation{%
  \institution{Houmo AI}
  \city{Nanjing}
  \country{China}
}

\author[X.~Si]{Xin Si}
\orcid{0000-0002-4993-0087}
\affiliation{%
  \institution{School of Integrated Circuits, Southeast University}
  \city{Nanjing}
  \country{China}
}

\author[J.~Yang]{Jun Yang}
\orcid{0000-0002-8379-0321}
\affiliation{%
  \institution{School of Integrated Circuits, Southeast University}
  \city{Nanjing}
  \country{China}
}

\author[G.~Sun]{Guangyu Sun}
\orcid{0000-0002-7315-6589}
\thanks{$^{\ddagger}$ Corresponding author, email: \href{mailto:gsun@pku.edu.cn}{gsun@pku.edu.cn}}
\affiliation{%
  \institution{School of Integrated Circuits, Peking University}
  \city{Beijing}
  \country{China}
}
\affiliation{%
  \institution{Beijing Advanced Innovation Center for Integrated Circuits}
  \city{Beijing}
  \country{China}
}

\renewcommand{\shortauthors}{C.~Xue, Y.~Wang, A.~Guo, Y.~Shi, J.~Zhou, X.~Dong, Y.~Yin, Y.~Zhang, T.~Jia, W.~Gao, Q.~Wu, X.~Si, J.~Yang, and G.~Sun}

\begin{abstract}
SRAM-based compute-in-memory (CIM) offers high computational density and energy efficiency for deep neural network (DNN) accelerators, but its limited capacity causes on/off-chip data movement overhead for large DNN models.
Existing CIM accelerator studies typically assume that DNN models fit entirely on-chip, leaving efficient dataflow design largely untapped. This paper introduces \papername{}, a systematic dataflow exploration framework for SRAM CIM accelerator, which addresses two key limitations of prior work. (1) It formulates a systematic dataflow design space spanning CIM macro configurations and macro-array organizations. (2) It introduces rigorous design evaluation using cycle-accurate architectural simulation and post-layout PPA analysis. We conduct an extensive design space exploration and apply \papername{} to representative LLM applications, providing practical insights for the principled design of CIM accelerators.
\end{abstract}



\maketitle

\section{Introduction}
\label{sec:introduction}

SRAM-based Compute-in-Memory (CIM) has emerged as a highly promising technique for accelerating deep neural networks (DNNs). By tightly integrating multiply-accumulate (MAC) units within the SRAM bitcell array, SRAM CIM provides superior computational density and energy efficiency over the disaggregated counterpart. These advantages position SRAM CIM as a competitive building block to implement matrix units within DNN accelerators. The immense potential of CIM has motivated not only extensive academic research~\cite{zhu2025leveraging,zhu2023mnsim,lee2024neurosim,andrulis2024cimloop,qi2025cimflow,wu2025exploiting,chen2023autodcim,zhang2024arctic,diao2025sega,shao2025syndcim,wang2025damil,sun2023analog}, but also commercial adoption across diverse application domains, such as d-Matrix for large language model serving~\cite{dmatrix}, Houmo for autonomous driving~\cite{houmo}, and MediaTek for smart mobile devices~\cite{DCIM-MediaTek}.

The broad application scope of SRAM CIM accelerators makes dataflow design critical to performance.
Prior CIM accelerator studies typically target DNN models that can fit entirely in the on-chip memory~\cite{zhu2023mnsim,lee2024neurosim,andrulis2024cimloop}. However, as DNN model sizes continue to scale, many contemporary models can no longer be fully stored on-chip. For instance, d-Matrix Corsair~\cite{dmatrix} integrates 2 GB of SRAM CIM, yet it cannot host the smallest member of the LLaMA-3 family with 8 billion parameters~\cite{dubey2024llama}.
Consequently, SRAM CIM accelerators for large DNN models must effectively manage on-/off-chip data movement. At the architecture level, weight updates and computation must be carefully scheduled to maximize performance and energy efficiency.
At the circuit level, the CIM macros and their interconnection must be co-designed to minimize integration overheads and ensure scalability.

\begin{table}[!t]
\caption{Comparison with existing researches related to design automation of SRAM CIM accelerators.}
\vspace{-10pt}
\centering
\scriptsize
\setlength{\tabcolsep}{1.4pt}
\resizebox{\columnwidth}{!}{%
\begin{tabular}{@{}
    >{\centering\arraybackslash}p{0.20\columnwidth}|
    >{\centering\arraybackslash}p{0.125\columnwidth}|
    >{\centering\arraybackslash}p{0.155\columnwidth}|
    >{\centering\arraybackslash}p{0.105\columnwidth}|
    >{\centering\arraybackslash}p{0.15\columnwidth}|
    >{\centering\arraybackslash}p{0.11\columnwidth}|
    >{\centering\arraybackslash}p{0.11\columnwidth}
@{}
}
\toprule
\makecell[c]{\textbf{Work}} & \makecell[c]{\textbf{Array}\\\textbf{Dataflow}} & \makecell[c]{\textbf{Array}\\\textbf{Interconnect}} & \makecell[c]{\textbf{Macro}\\\textbf{Capacity}} & \makecell[c]{\textbf{Compute-I/O}\\\textbf{Overlap}} & \makecell[c]{\textbf{Macro}\\\textbf{PPA}} & \makecell[c]{\textbf{Array}\\\textbf{PPA}} \\ \midrule \midrule
\makecell[c]{AutoDCIM et al.\\\cite{chen2023autodcim,zhang2024arctic,diao2025sega,shao2025syndcim}} & --- & --- & Flexible & Unsupported & \makecell[c]{Post\\Layout} & --- \\ \midrule
\makecell[c]{CiMLoop et al.\\\cite{zhu2023mnsim,lee2024neurosim,andrulis2024cimloop,qi2025cimflow,wu2025exploiting}} & \makecell[c]{Weight\\Stationary} & Broadcast & Flexible & Supported & \makecell[c]{Analytical\\Model} & \makecell[c]{Analytical\\Model} \\ \midrule
\makecell[c]{CIM-MXU\\\cite{zhu2025leveraging}} & \makecell[c]{Output\\Stationary} & Systolic & Fixed & Supported & \makecell[c]{Post\\Layout} & \makecell[c]{Analytical\\Model} \\ \midrule
\makecell[c]{\textbf{\papername{}}\\\textbf{(Ours)}} & \makecell[c]{\textbf{WS \& OS}} & \makecell[c]{\textbf{Broadcast}\\\textbf{\& Systolic}} & \textbf{Flexible} & \makecell[c]{\textbf{Unsupported}\\\textbf{\& Supported}} & \makecell[c]{\textbf{Post}\\\textbf{Layout}} & \makecell[c]{\textbf{Post}\\\textbf{Layout}} \\ \bottomrule
\end{tabular}%
}
\label{tab:review}
\vspace{-15pt}
\end{table}

\begin{figure}[!t]
    \centering
    \includegraphics[width=0.9\linewidth]{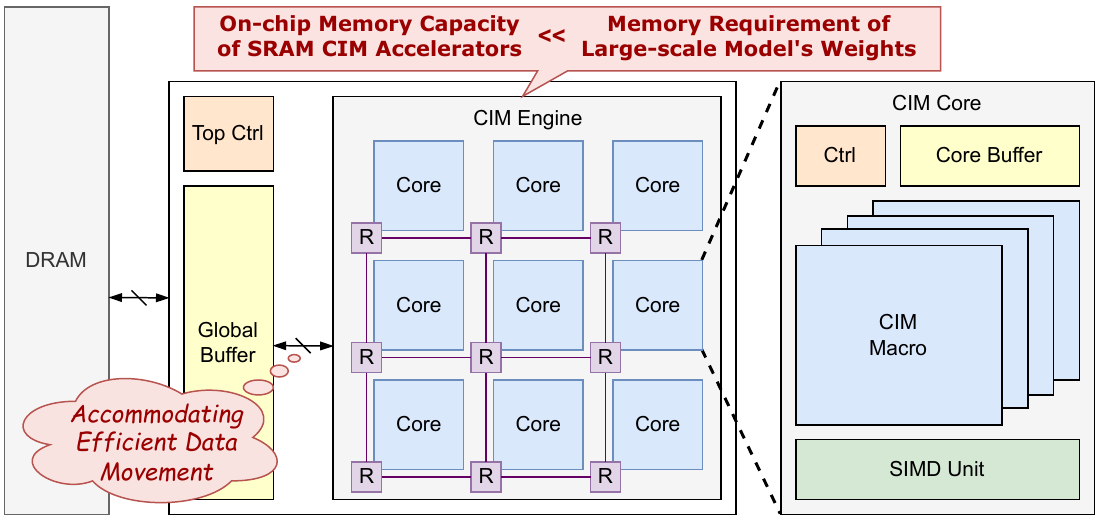}
    \caption{The design hierarchy of typical CIM accelerators. The limited capacity of SRAM CIM renders dataflow design critical to accommodate efficient data movement.}
    \label{fig:arch_overview}
    \vspace{-10pt}
\end{figure}

Despite numerous studies on CIM accelerator design and automation, these works cannot adequately address the aforementioned emerging requirements. Table~\ref{tab:review} summarizes relevant studies and their approaches to design space construction and design evaluation of CIM accelerators, from which we identify the following limitations:
\textbf{(1) Limited exploration of macro array organization.} Existing architectural modeling frameworks predominantly adopt weight-stationary dataflow with input-broadcast interconnects~\cite{zhu2023mnsim,lee2024neurosim,andrulis2024cimloop,qi2025cimflow,wu2025exploiting}, without exploring alternatives such as output-stationary dataflow~\cite{zhu2025leveraging} and systolic interconnection~\cite{zhu2025leveraging,chen2023systolic,zhan2024gslp}. Therefore, a principled comparison of CIM macro spatial arrangements remains missing. 
\textbf{(2) Simplified assumption of CIM macro capabilities.} Most frameworks assume that their CIM macros support simultaneous computation and I/O access, referred to as compute-I/O overlap~\cite{fujiwara20225,fujiwara202434}. However, this functionality involves a system-level design trade-off between reduced end-to-end latency and physical implementation overhead, which is neglected by these frameworks. 
\textbf{(3) Inaccurate design evaluation of CIM macro and spatial arrangements.} First, analytical models used to evaluate different CIM macros often have limited fidelity, since they are calibrated against a very limited set of silicon prototypes~\cite{sun2023analog,lee2024neurosim}. Moreover, array-level physical design impacts are largely overlooked. Placement and routing introduce nontrivial integration overheads to the area footprint. Consequently, such inaccurate assessment may lead to sub-optimal CIM accelerator design.

To bridge these research gaps, this paper proposes a systematic dataflow exploration framework for SRAM CIM accelerators, namely \papername{}. By establishing a comprehensive dataflow design space and systematically exploring the design tradeoffs of end-to-end system quality-of-results (QoRs), \papername{} can automatically identify the Pareto-optimal CIM accelerator dataflow designs for the target DNN models.
The main contributions of this paper are summarized as follows:
\begin{itemize}[leftmargin=15pt, labelindent=0pt]
    \item We formulate a systematic CIM accelerator dataflow design space, encompassing both spatial arrangements of macro array and macro design options.
    \item We construct a rigorous CIM accelerator design evaluator to accurately model both architectural performance and post-layout implementation cost.
    \item We conduct design exploration on representative DNN applications and provide insightful takeaways.
\end{itemize}

The remainder of this paper is organized as follows:
Section~\ref{sec:background} introduces the background and motivation.
Section~\ref{sec:methodology} introduces \papername{}'s framework.
Section~\ref{sec:evaluation} introduces the evaluation results.
Finally, section~\ref{sec:conclusion} concludes this paper.

\section{Background and Motivation}
\label{sec:background}

In this section, we introduce the background of CIM accelerators and explain our motivation for systematic dataflow exploration.

\begin{figure}[t]
    \centering
    \begin{subfigure}{0.49\linewidth}
        \centering
        \includegraphics[width=\linewidth]{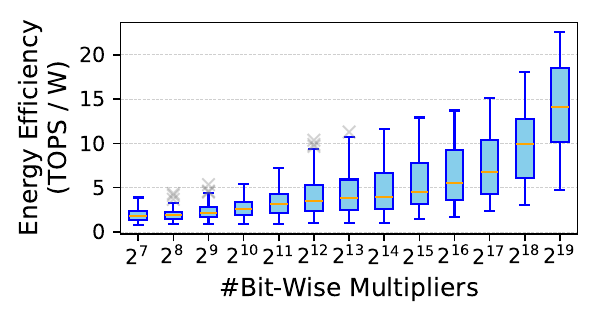}
    \end{subfigure}
    \hfill
    \begin{subfigure}{0.49\linewidth}
        \centering
        \includegraphics[width=\linewidth]{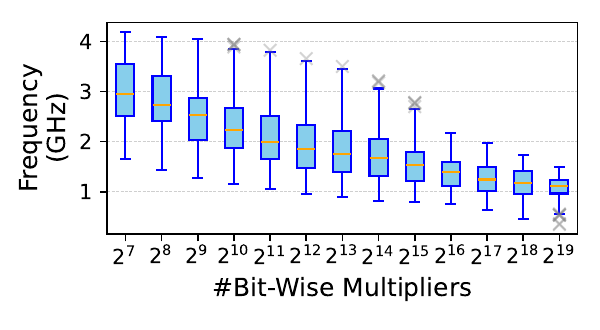}
    \end{subfigure}
    
    \vspace{-10pt}

    \caption{Distribution of CIM macro energy efficiency and frequency under different compute capacities.}
    \label{fig:macro_tradeoff}
    \vspace{-10pt}
\end{figure}

\begin{figure}[t]
    \centering
    \begin{subfigure}{0.49\linewidth}
        \centering
        \includegraphics[width=\linewidth]{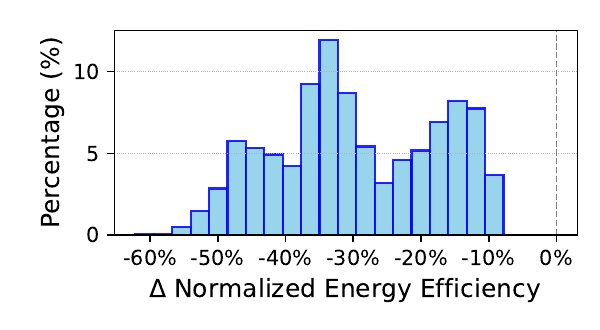}
    \end{subfigure}
    \hfill
    \begin{subfigure}{0.49\linewidth}
        \centering
        \includegraphics[width=\linewidth]{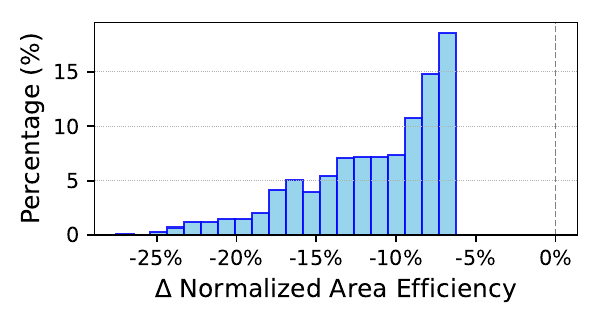}
    \end{subfigure}
    
    \vspace{-10pt}

    \caption{Histogram of CIM macro energy/area efficiency relative degradation when compute-I/O overlap is enabled.}
    \label{fig:overlap_tradeoff}
    \vspace{-10pt}
\end{figure}

\subsection{CIM Accelerator Design}

As shown in Figure~\ref{fig:arch_overview}, SRAM CIM accelerators consist of multiple design levels.
The \textit{CIM macros} are the tight integration of SRAM cell arrays and multiply-accumulate units, typically made by manual layout design or custom automated design tools~\cite{chen2023autodcim,zhang2024arctic,diao2025sega,shao2025syndcim}.
Multiple CIM macros are interconnected with a fixed hardware dataflow to form a \textit{CIM core}, and multiple cores are further connected via a flexible network-on-chip (NoC) to construct a \textit{CIM engine}.

The design options across CIM accelerator hierarchy constitute an enormous design space, motivating a series of design space exploration (DSE) studies~\cite{zhu2023mnsim,lee2024neurosim,andrulis2024cimloop,qi2025cimflow,wu2025exploiting}. However, these works give limited attention to design trade-offs for CIM core organization, which motivates us to conduct a more systematic exploration: 
(1) \underline{\textit{Weight-stationary dataflow}} is widely adopted by existing studies, since the targeted DNN models typically fit entirely within on-chip memory. In this setting, weight swapping overhead is often treated as a one-time cost or negligible relative to computation. However, this conclusion may break down for large DNN models that require weight swapping between on-/off-chip memory.
(2) \underline{\textit{Input-broadcast interconnect}} is also commonly adopted for its simplicity. However, its physical implementation is challenging since global wires between core buffer and CIM macros take up substantial routing resources, rendering its poor scalability.

\begin{figure*}[!t]
    \centering
    \begin{minipage}[t]{0.32\textwidth}
        \centering
        \begin{minipage}[c][0.17\textheight][c]{\linewidth}
            \centering
            \small
            \captionof{table}{\papername{}'s design space}
            \vspace{-10pt}
            \setlength{\tabcolsep}{2pt}
            \renewcommand{\arraystretch}{0.95}
            \resizebox{\linewidth}{!}{%
            \begin{tabular}{c|c|c}
            \toprule
            \textbf{Hierarchy}     & \textbf{Parameter}           & \textbf{Candidate Values} \\ \midrule
            \multirow{7}{*}{Macro} & Accumulation Length ($AL$)   & 8,16,32,...,256     \\
                                   & Local Storage Length ($LSL$) & 2,4,8,...,64        \\
                                   & Parallel Channel ($PC$)      & 2,4,8,...,256       \\
                                   & Pipeline Level ($PL$)        & 0,1,2,3,4,5         \\
                                   & Compute-I/O Overlap ($OL$)   & True, False         \\
                                   & Weight Bitwidth ($WBW$)      & 8                   \\
                                   & Input Bitwidth ($IBW$)       & 8                   \\ \midrule
            \multirow{5}{*}{Array} & Array Row Size ($BR$)        & 1,2,3,...,64        \\
                                   & Array Column Size ($BC$)     & 1,2,3,...,64        \\
                                   & Dataflow                     & WS, OS              \\
                                   & Interconnection              & Broadcast, Systolic \\ \bottomrule
            \end{tabular}%
            }
            \label{tab:design_space}
        \end{minipage}

    \end{minipage}
    \hfill
    \begin{minipage}[t]{0.32\textwidth}
        \centering
        \begin{minipage}[c][0.17\textheight][c]{\linewidth}
            \centering
            \includegraphics[width=\linewidth,height=0.17\textheight,keepaspectratio]{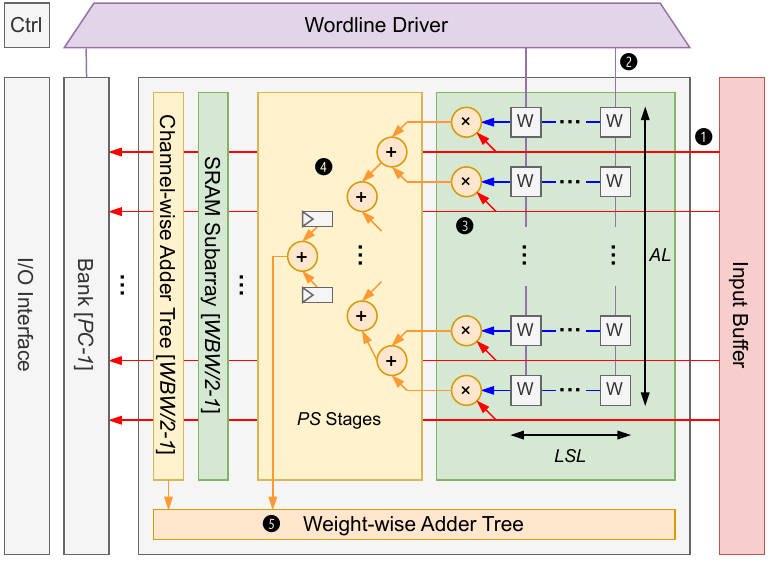}
        \end{minipage}

        \begin{minipage}[t]{\linewidth}
            \captionsetup{skip=2pt}
            \captionof{figure}{\papername{} CIM macro template.}
            \label{fig:macro}
        \end{minipage}
    \end{minipage}
    \hfill
    \begin{minipage}[t]{0.32\textwidth}
        \centering
        \begin{minipage}[c][0.17\textheight][c]{\linewidth}
            \centering
            \includegraphics[width=\linewidth,height=0.17\textheight,keepaspectratio]{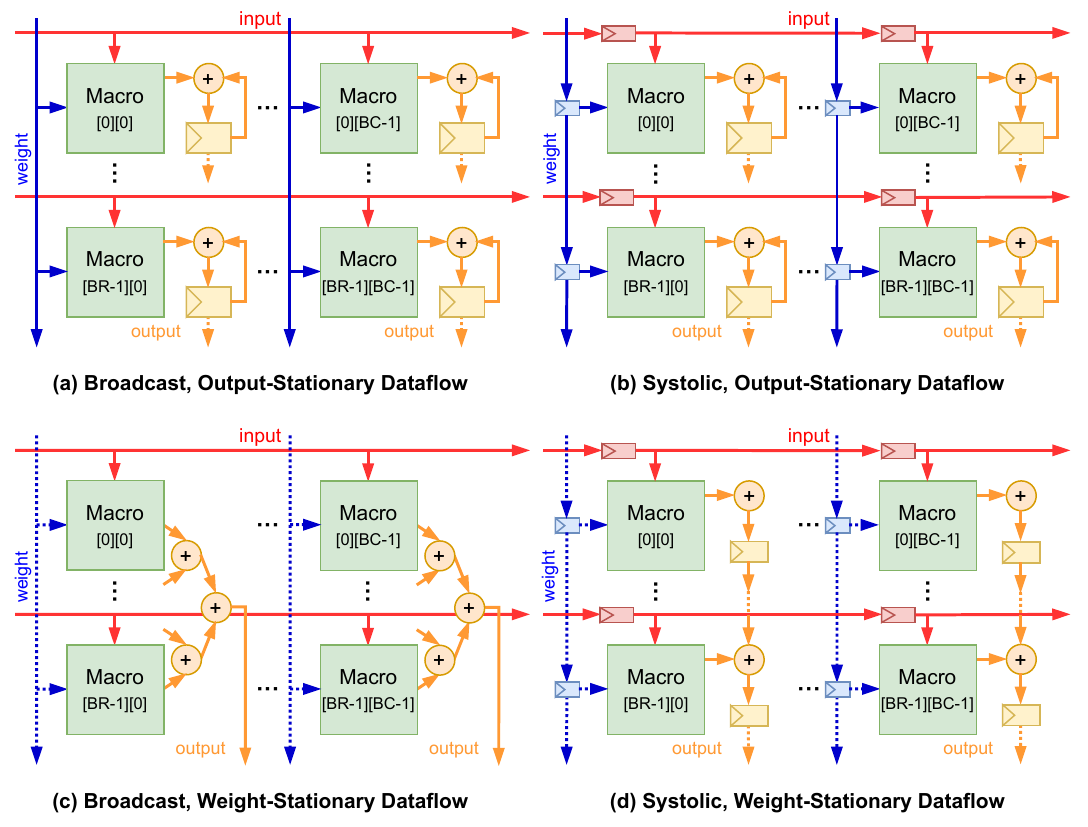}
        \end{minipage}

        \begin{minipage}[t]{\linewidth}
            \captionsetup{skip=2pt}
            \captionof{figure}{\papername{} dataflow.}
            \label{fig:dataflow}
        \end{minipage}
    \end{minipage}
\vspace{-10pt}
\end{figure*}

\subsection{CIM Macro Design}

Another aspect of our motivation originates from the post-layout evaluation of CIM macro design variants.
We leverage an open-source CIM macro compiler~\cite{cimcompiler} to produce the results in Figure~\ref{fig:macro_tradeoff} and \ref{fig:overlap_tradeoff}.
Our analysis reveals two CIM macro design trade-offs largely neglected by previous CIM accelerator DSE research:
(1) \underline{\textit{Macro capacity}} introduces the design trade-off between energy efficiency and performance, as shown in Figure~\ref{fig:macro_tradeoff}. Larger CIM macros with more bit-wise multipliers (i.e. higher \textit{compute capacity}) exhibit better energy efficiency due to customized compute–memory integration, while smaller macros with fewer multipliers and shorter timing critical paths can operate at higher frequencies. Unfortunately, these physical characteristics are not properly modeled in prior system-level exploration, which potentially leads to sub-optimal designs. 
(2) \underline{\textit{Compute-I/O Overlap}} minimizes weight swapping overhead but increases power consumption and silicon area, as shown in Figure~\ref{fig:overlap_tradeoff}. Despite this trade-off, prior studies typically assume this feature to be enabled by default, with limited assessment of the system-level impacts.

\section{\papername{} Framework}
\label{sec:methodology}

In this section, we first introduce \papername{}'s dataflow design space in Section~\ref{sec:architecture}. Then, we detail the architectural modeling of dataflow variants in Section~\ref{sec:dataflow}. Finally, we introduce the post-layout design evaluation flow in Section~\ref{sec:generator}.

\subsection{\papername{} Architecture}
\label{sec:architecture}

To analyze the trade-offs between performance, area, and power in \papername{}, we consider the joint design space of CIM macros as well as the macro array arrangement. Table~\ref{tab:design_space} summarizes the related design parameters and their candidate values.


\noindent \textbf{Macro.}
\papername{} adopts a CIM macro design template similar to~\cite{guo202428}, as shown in Figure~\ref{fig:macro}.
The CIM macro contains $PC$ banks, each storing a weight block consisting of $LSL$ weight rows and $AL$ weight columns.
Within each bank, the weight block of bitwidth $WBW$ is sliced into 2-bit chunks and stored distributively in multiple subarrays.
Each subarray is augmented with peripheral circuits for MAC operations, including bit-wise multipliers and adder trees.
The I/O interface is dedicated to weight updates. Notably, previous work typically assumes that the CIM macros support simultaneous computation and weight updates by default~\cite{zhu2025leveraging,wu2025exploiting}. While this feature reduces the outstanding latency for weight updates, it also necessitates additional hardware resources (e.g. bitlines and wordline drivers). Therefore, \papername{} treats compute-I/O overlap as an optional feature and explores its related design trade-offs.  

The CIM macro computes general matrix-vector multiplication (GEMV) as follows:
For an input vector of bitwidth $IBW$, the input buffer broadcasts two of its bit-slices to all SRAM subarrays (Step \blackcircle{1}), while the wordline driver simultaneously activates one row within each SRAM subarray (Step \blackcircle{2}).
This triggers peripheral bit-wise multipliers to compute the products in parallel (Step \blackcircle{3}).
The multiplication results are first sent to the subarray-level adder tree for channel-wise reduction (Step \blackcircle{4}), and then sent to the bank-level adder tree for weight-wise reduction (Step \blackcircle{5}). The reduction takes $PS$ cycles to complete, as combinatorial logic has been pipelined to shorten the critical path and improve the operating frequency of the CIM macro. 
Ignoring the fill and drain latencies of pipelined accumulation logic, the computational behavior of CIM macros can be summarized as follows: \textit{Each CIM macro computes $PC$ parallel dot products between the input activation vector and $PC$ weight rows every $IBW/2$ cycles.} 


\subsection{\papername{} Dataflow}
\label{sec:dataflow}

In this section, we discuss how the parameters in Table~\ref{tab:design_space} govern the partitioning and scheduling of the global matrix multiplication for the four array-level dataflows illustrated in Figure~\ref{fig:dataflow}.


\noindent \textbf{Macro-level GEMM.} A single CIM macro stores a weight block of size $(PC \times LSL) \times AL$, and the activation block size is $AL \times TL$. A \textit{weight row} refers to the weights at the same row index across all $PC$ banks. The $PC$ banks perform parallel computations, with serial weight updates. The cycle count for processing one activation block with one weight row is 
\begin{equation}
T_c=TL \cdot IBW/2,
\end{equation}
while the cycle count for updating one weight row is 
\begin{equation}
T_s=\kappa \cdot PC \cdot WBW ,    
\end{equation}
where $\kappa$ represents the intrinsic weight-write speed of the CIM macros.

For multiplication within a single weight and activation block, the outer loop iterates over the $LSL$ weight rows, and the inner loop iterates over the $TL$ activation columns, keeping the same wordline active for multiple inputs. By minimizing wordline toggling, this approach reduces switching power in CIM macros under the same area and performance constraints. Once a macro completes the multiplication in the current weight row, it initiates the weight update for this row and proceeds to the next weight row. The activation block is reused to multiply with the subsequent weight row until all $LSL$ weight rows are processed. All dataflows follow this loop nesting; they differ mainly in how activations and weights move across the array during each block matrix multiplication.


\noindent \textbf{Array-level GEMM.}
For \textit{WS-Broadcast} dataflow, the reduction tree requires macros to start and end computation simultaneously. After macros complete the computation of the current weight row, weight updates occur in a row-by-row manner in each column, where macros receive updates sequentially. Since the next-round activation block must be delivered concurrently to all macros, while one macro updates its weight, the others in the column are idle, neither updating weights nor computing.

In contrast, \textit{WS-Systolic} dataflow avoids this issue by staggering the entry of activation blocks horizontally across rows with a $\kappa \cdot PC \cdot WBW$ cycle delay. Once the activations complete the computation of the current weight row, new weights enter each column in a similar staggered manner. This ensures that each macro can update the current weight row upon finishing computation, enabling all macros to perform at least one of the tasks between computation and weight updates.

For the \textit{OS-Broadcast} dataflow, activation movement follows the WS-Broadcast pattern, but weight updates differ: macros within the same column share a common weight block, and the updated weights are broadcast synchronously to all macros in the column.

In the \textit{OS-Systolic} flow, once a macro completes local computation, it begins updating its current weight row with the weight passed from its upstream neighbor. When the weight update finishes, the macro receives the initial bits of next-round activation block from its left neighbor, which will be used for computation in the current cycle.

\begin{figure}[!t]
    \centering
    \includegraphics[width=1\linewidth]{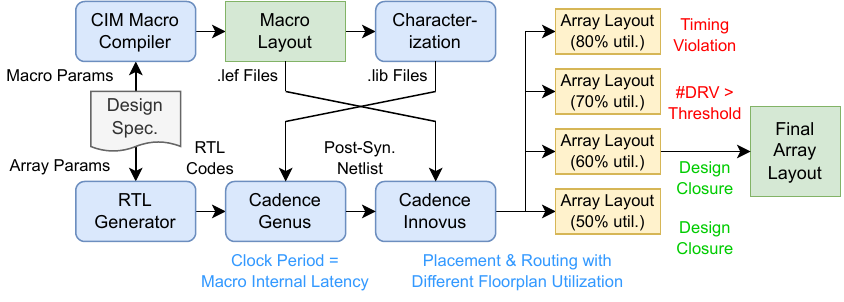}
    \vspace{-20pt}
    \caption{\papername{}'s macro array generator workflow.}
    \label{fig:evaluation_flow}
    \vspace{-10pt}
\end{figure}

\begin{figure}[!t]
    \centering
    \includegraphics[width=0.8\linewidth]{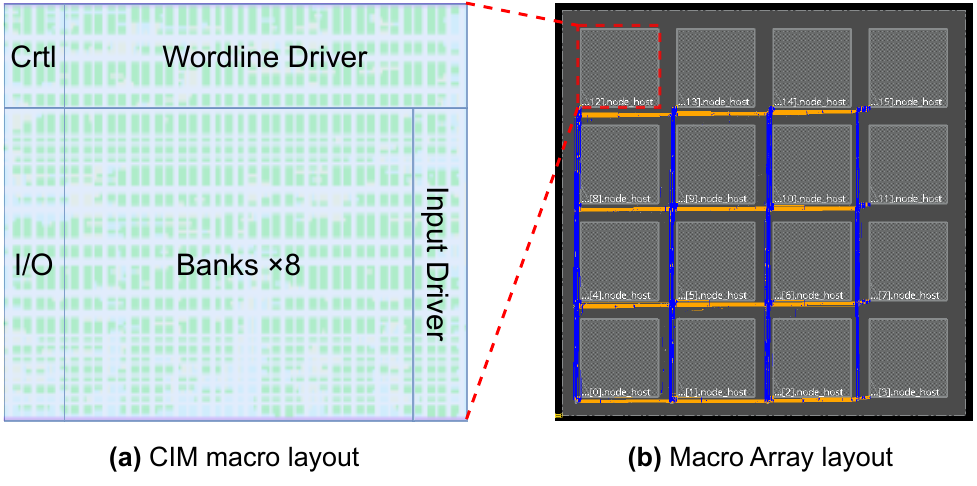}
    \vspace{-10pt}
    \caption{Example layouts from \papername{}'s macro array generator.}
    \label{fig:layout}
    \vspace{-10pt}
\end{figure}

\noindent \textbf{Compute-I/O Overlap.} If a macro supports compute-I/O overlap, when a macro completes the multiplication on the current weight row, the computation on the next row can begin concurrently with the update of the current row. Otherwise, computation must wait until the weight update is completed.

In \textit{WS-Broadcast} dataflows, all macros perform computations synchronously, requiring row-by-row weight updates across the array. When compute-I/O overlap is supported, computation on the next weight row proceeds with the same activation block during the update. If the condition $BR \cdot T_s \leq T_c$ is met, all macros' weights are updated before the next round of activation blocks is delivered, thus addressing the issue of idle macros effectively. 

For other dataflows, enabling compute-I/O overlap reduces latency, but the overall flow remains largely unchanged. The primary effect is a reduction in latency for each macro's computation and weight update. If compute-I/O overlap is not supported, the cycle count for multiplication between a single weight block and activation block is given by:
\begin{equation}
T_{nol}=LSL\times (T_s+T_c) .
\end{equation}
Compute-I/O overlap reduces block matrix multiplication cycles to:
\begin{equation}
T_{ol}=LSL\times \max(T_s,T_c) .
\end{equation}
The optimization ratio for cycle count is:
\begin{equation}
1-\frac{T_{ol}}{T_{nol}}=1-\frac{\max(T_s,T_c)}{T_s+T_c}\leq0.5 .
\end{equation}
Thus, enabling compute-I/O overlap can reduce latency by up to 50\%. However, among the best-performing designs in most DNN accelerators, $T_c$ is typically several times larger than $T_s$ due to the large computational density of SRAM CIM, so the actual latency reduction is often modest.

\subsection{CIM Macro Array Generator}
\label{sec:generator}

To facilitate an accurate assessment of the post-layout quality for various designs, we developed an automatic generator to produce CIM macro array layout, whose workflow is shown in Figure~\ref{fig:evaluation_flow}.

The workflow starts with CIM macro layout generation. Given the CIM macro specification, we employ an open source SRAM CIM compiler to generate the corresponding macro layout~\cite{cimcompiler}. Examples of the generated CIM macro layout are shown in Figure~\ref{fig:layout}(a). The CIM macro layout undergoes SPICE simulation to extract its timing and power characteristics, which are then compiled into a standard timing library \texttt{.lib} file. Concurrently, the macro's geometric information is abstracted into a library exchange format \texttt{.lef} file. The \texttt{.lib} and \texttt{.lef} files will be leveraged in the subsequent digital design flow.

Next, we generate the macro array layout. We first translate the macro array specification into Register-Transfer Level (RTL) codes. Subsequently, we use Cadence Genus to synthesize the RTL codes into gate-level netlists, with the target operational frequency determined by the internal delay of the CIM macro. Then, we use Cadence Innovus for placement and routing. An example of the post-routing macro array layout is shown in Figure~\ref{fig:layout}(b). To evaluate routing resource requirements of different interconnection topologies, the P\&R process is executed on floorplans with varying utilization rates. The final floorplan of the CIM macro array is the most compact implementation, which satisfies the timing constraints while maintaining an acceptable amount of design rule violations (DRVs). Based on the final floorplan, we can obtain its area and the power overhead for array integration.

\section{Evaluation}
\label{sec:evaluation}

\subsection{Experimental Setup}



\papername{} explores the dataflow design space described in TABLE~\ref{tab:design_space}, and employs Bayesian optimization to identify Pareto-optimal designs~\cite{li2021openbox}.
We developed an in-house cycle-accurate simulator to obtain the total cycles for running specific applications.
All CIM macro arrays are equipped with an identical core buffer, which provides sufficient capacity and bandwidth to fully utilize all the computational resources.
The physical layouts for the CIM macros and the full array are generated using a 28nm process technology.
For all of the designs explored, we evaluate their performance, power, and area (PPA) defined as follows: 
\begin{itemize}[nosep, leftmargin=15pt, labelindent=0pt]
    \item \textit{Performance}: In the absence of a specific application, we quantify CIM macro array performance by the theoretical peak throughput, i.e. the number of MAC units times operational frequency. When a specific application is considered, we use end-to-end latency as the performance metric.
    \item \textit{Power}: We first conduct post-layout analysis to extract the static and dynamic power consumption of the CIM macros and other components within the array. Then we derive the transition rates from cycle-accurate simulation traces. Finally, we integrate these results to calculate the overall power consumption.
    \item \textit{Area}: The total area is extracted from the CIM macro array floorplan that has the highest utilization while achieving our design closure standard described in Section~\ref{sec:generator}.
\end{itemize}






\begin{figure}[t]
    \centering
    \vspace{0pt}
    \captionsetup[subfigure]{skip=0pt}

    \begin{subfigure}{\linewidth}
        \centering
        \includegraphics[width=0.9\linewidth]{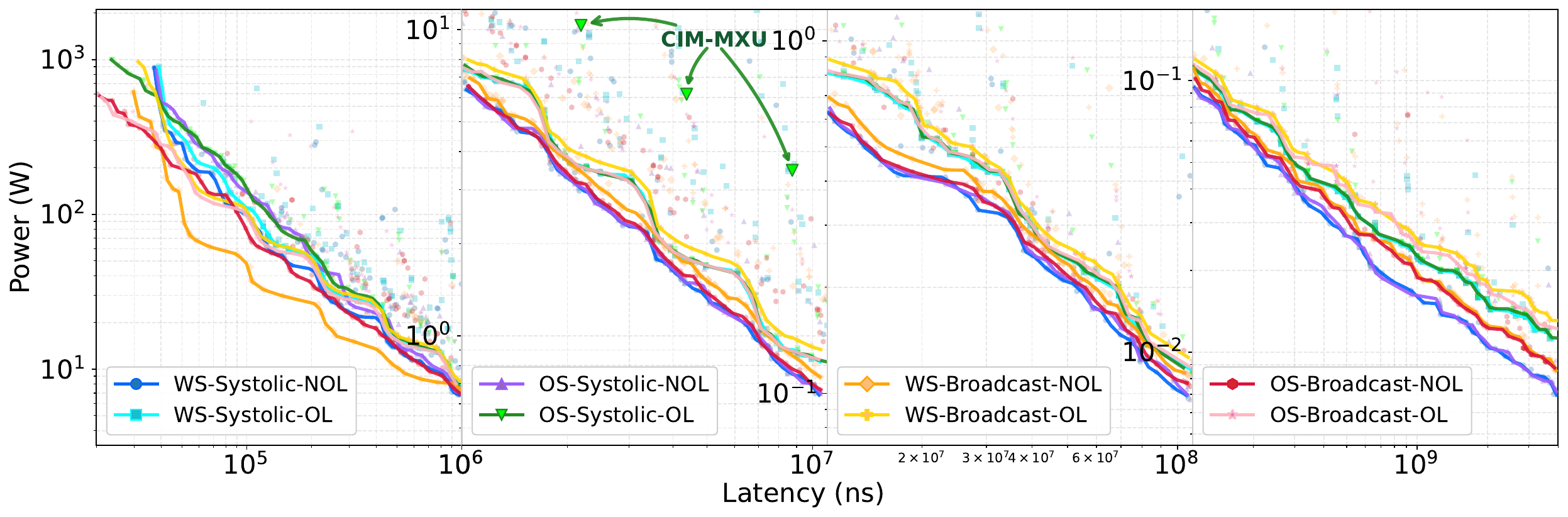}
        \caption{Performance and power}
    \end{subfigure}

    \vspace{2pt}

    \begin{subfigure}{\linewidth}
        \centering
        \includegraphics[width=0.9\linewidth]{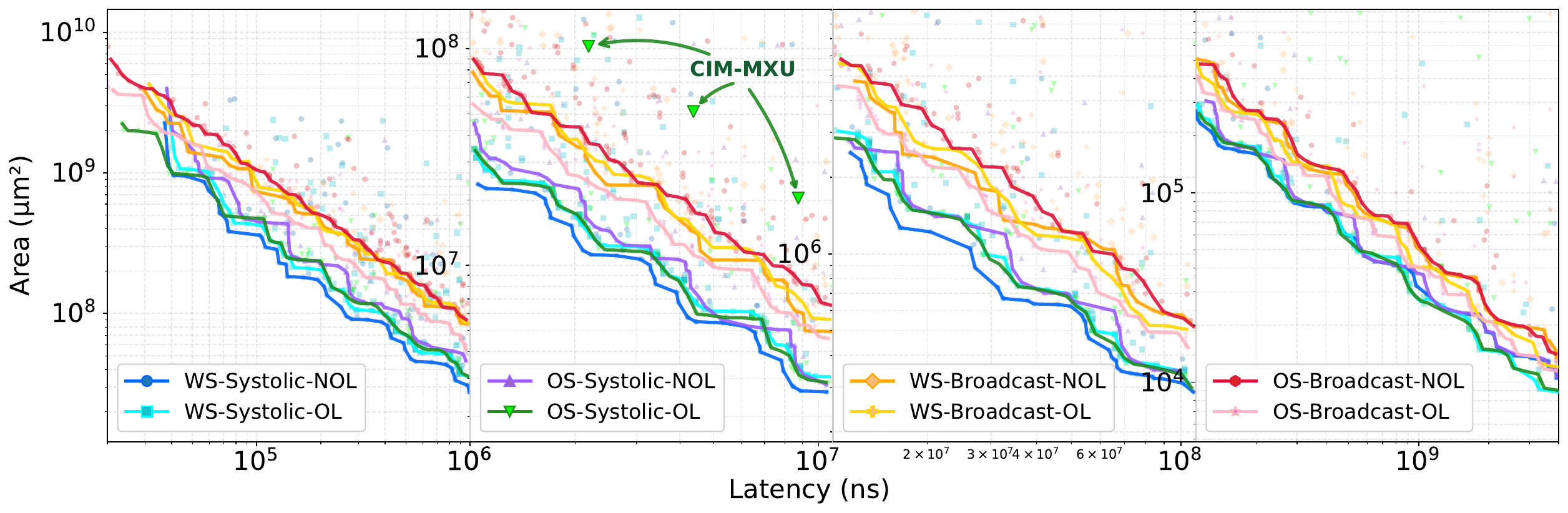}
        \caption{Performance and area}
    \end{subfigure}

    \vspace{-10pt}
    \caption{Pareto frontiers of different dataflows in \papername{}.}
    \label{fig:pareto}
    \vspace{-10pt}
\end{figure}


\begin{table*}[!t]
\caption{Optimized design points for different large language model inference tasks}
\vspace{-6pt}
\resizebox{\linewidth}{!}{%
\begin{tabular}{c|c|c|c|c|c|c|c|c|c}
\toprule
\textbf{Model}      & \textbf{\#Layer} & \textbf{Hidden Dim} & \textbf{Sequence Length} & \textbf{\#CIM Core} & \textbf{Best Dataflow}   & \textbf{(LSL,AL,PC,PL,BC,BR,TL)} & \textbf{Latency(ms)} & \textbf{Power(W)} & \textbf{Area(mm$^2$)} \\ \midrule
Qwen3-0.6B & 28      & 1024       & 8192            & 1          & OS-Systolic-OL  & (2,128,8,2,12,2,16)     & 216.552     & 1.869    & 2.916  \\ 
Llama3-8B  & 32      & 4096       & 8192            & 4          & OS-Systolic-OL  & (2,256,16,4,4,2,32)     & 886.272     & 0.994    & 2.824  \\ 
Llama3-70B & 80      & 8192       & 8192            & 8          & OS-Systolic-OL  & (2,256,16,4,4,2,32)     & 5771.04     & 0.996    & 2.839  \\ 
GPT3-175B  & 96      & 12288      & 2048            & 16         & WS-Systolic-NOL & (2,256,4,2,3,8,256)     & 2221.92     & 0.842    & 1.993  \\ 
GPT3-175B  & 96      & 12288      & 131072          & 64         & WS-Systolic-NOL & (2,256,4,2,2,12,512)    & 35753.76    & 0.904    & 2.062  \\ \bottomrule     
\end{tabular}
}%
\label{tab:case_study}
\end{table*}

\subsection{Main Results}

In this section, we compare different CIM accelerator design candidates in large language model inference application. We select LLaMA-3-8B~\cite{dubey2024llama} as the target model and employ W8A8 quantization to enable integer-only GEMM. We assume that the CIM core handles a batch size of 8 requests, with each request having a sequence length of 1024 tokens. When focusing on Q/K/V projection operations, the corresponding GEMM sizes $M, N, K$ are 8192, 4096, 4096, respectively. The Pareto frontiers for \textit{performance–area} and \textit{performance–power} are illustrated in Figure~\ref{fig:pareto}. We also present the efficiency of the design candidates explored by previous work CIM-MXU~\cite{zhu2025leveraging}.



As shown in Figure~\ref{fig:pareto}(a), the dataflows differ substantially in terms of area efficiency. Systolic dataflows consistently outperform broadcast in area efficiency, owing to reduced requirements on routing resources. Within the systolic family, the \textit{output-stationary without overlap} (OS–Systolic–NOL) variant is suboptimal due to its limited reuse and lower utilization. Among the remaining three systolic flows (WS–NOL, WS–OL, and OS–OL), WS–NOL achieves the best area efficiency across nearly the entire performance range, with only extremely low-area regions exhibiting cases where WS–OL or OS–OL achieve slightly lower area overhead.
In contrast to significant area disparities, Figure~\ref{fig:pareto}(b) shows that the power differences between dataflows are relatively small. For dataflows that employ macros supporting compute–I/O overlap, they exhibit noticeably lower energy efficiency, primarily due to more frequent buffering and switching activities. WS-Broadcast-NOL offers better energy efficiency in the high-compute-capacity regions, whereas the two systolic flows without compute–I/O overlap (WS–Systolic–NOL and OS–Systolic–NOL) become more favorable across most of the remaining operating range, owing to their simpler control and reduced simultaneous data access.

\begin{figure}[t]
    \centering
    \centering
    \includegraphics[width=0.9\linewidth]{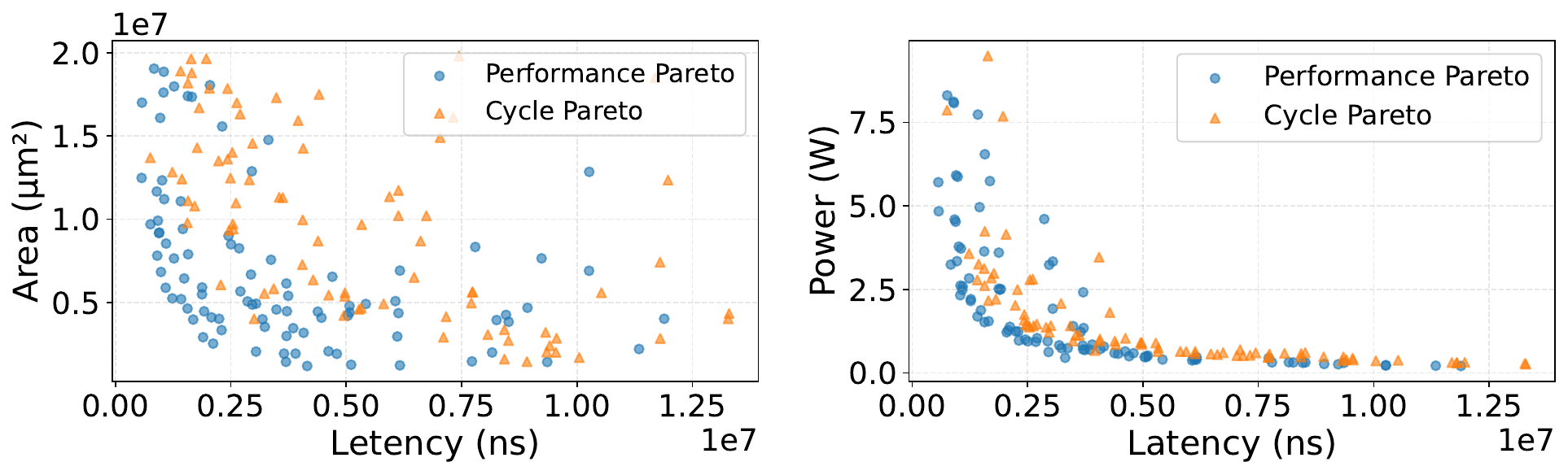}
    \vspace{-10pt}
    \caption{Comparison of cycle-oriented and performance-oriented design Pareto frontiers.}
    \label{fig:cycle_freq}
    \vspace{-10pt}
\end{figure}

\begin{figure}[t]
    \centering
    \begin{subfigure}{0.49\linewidth}
        \centering
        \includegraphics[width=\linewidth]{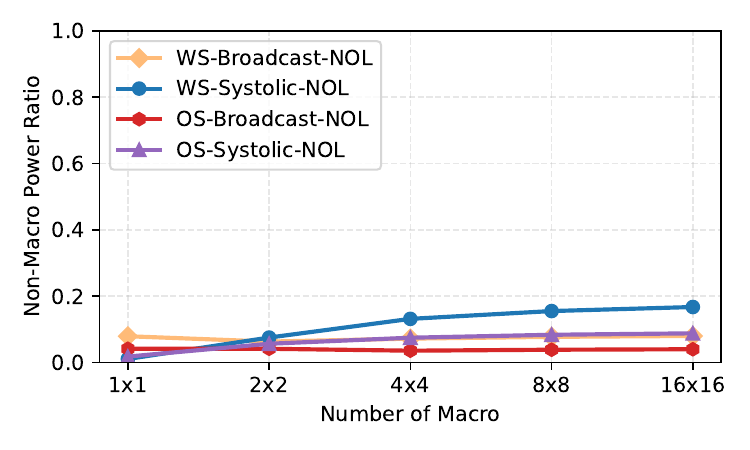}
        \caption{Power overhead}
    \end{subfigure}
    \begin{subfigure}{0.49\linewidth}
        \centering
        \includegraphics[width=\linewidth]{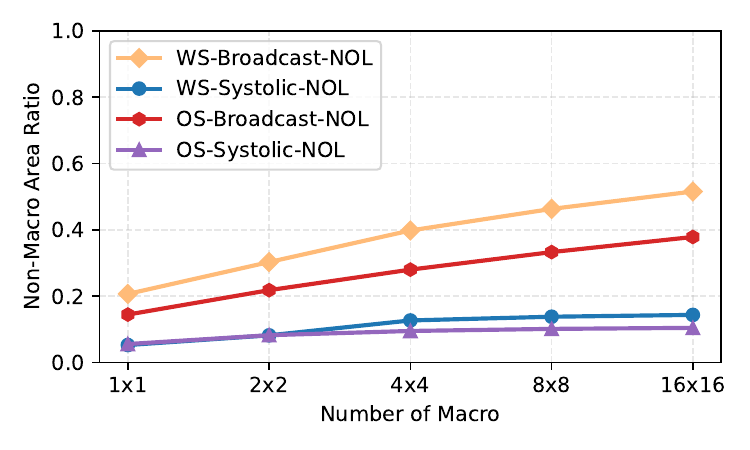}
        \caption{Area overhead}
    \end{subfigure}
    
    \vspace{-10pt}
    \caption{The contribution of non-macro components to overall power and area under different array sizes.}
    \label{fig:scalability}
    \vspace{-10pt}
\end{figure}

To claim the importance of macro–array co-design and accurate PPA evaluation in \papername{}, we take the WS–Systolic–NOL dataflow as an example. For compute throughput—the most scrutinized metric—performance is jointly determined by both frequency and total cycle count. If one focuses solely on cycles and performs array-level exploration without timing-aware modeling\cite{zhu2025leveraging}, the resulting Pareto front indeed appears more favorable in terms of cycles. However, once the impact of frequency is taken into account, the actual achievable performance becomes inferior to that predicted by such a cycle-only exploration. This discrepancy is evident in Figure~\ref{fig:cycle_freq}, demonstrating that an accurate and comprehensive evaluation framework is essential for truly optimal design points.

\subsection{Performance Analysis}

In this section, we conduct ablation studies on the system-level impact of individual design options, including array size, macro size, and compute-I/O overlap. Based on the evaluation results, we distill takeaways for future CIM accelerator designers.

\noindent \textbf{Array Organization.}
We evaluate the scalability of different array organizations by analyzing the overhead of integrating a different number of CIM macros. The same CIM macro with 4 TOPS computational capacity is adopted across all array configurations. Figure~\ref{fig:scalability}(a) shows that the power overhead for all the dataflows remains consistently below 20\%. However, Figure~\ref{fig:scalability}(b) shows that broadcast dataflows incur a much higher area overhead than systolic dataflows, since their global interconnections demand more routing resources than localized connections of systolic dataflows.

\takeaway{Relative to exclusive intra- and inter-macro broadcasting, integrating systolic interconnects can enhance area efficiency without degrading energy efficiency.}

\begin{figure}[t]
    \centering
    \captionsetup{skip=2pt}
    \begin{subfigure}{0.49\linewidth}
        \centering
        \includegraphics[width=\linewidth]{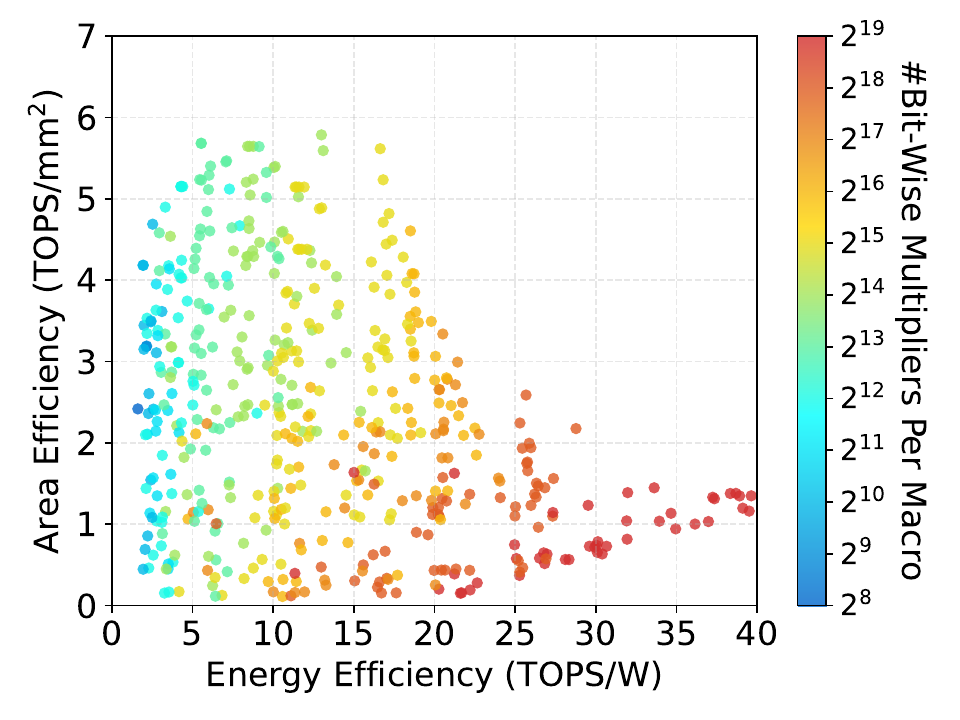}
        \caption{WS-Systolic}
    \end{subfigure}
    \begin{subfigure}{0.49\linewidth}
        \centering
        \includegraphics[width=\linewidth]{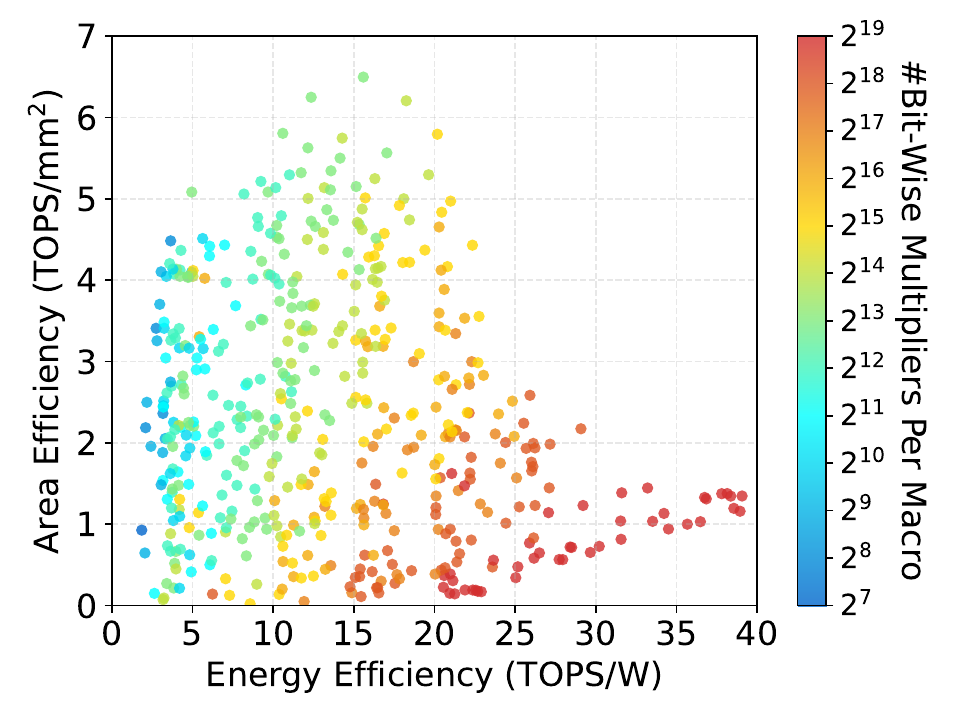}
        \caption{OS-Systolic}
    \end{subfigure}

    \begin{subfigure}{0.49\linewidth}
        \centering
        \includegraphics[width=\linewidth]{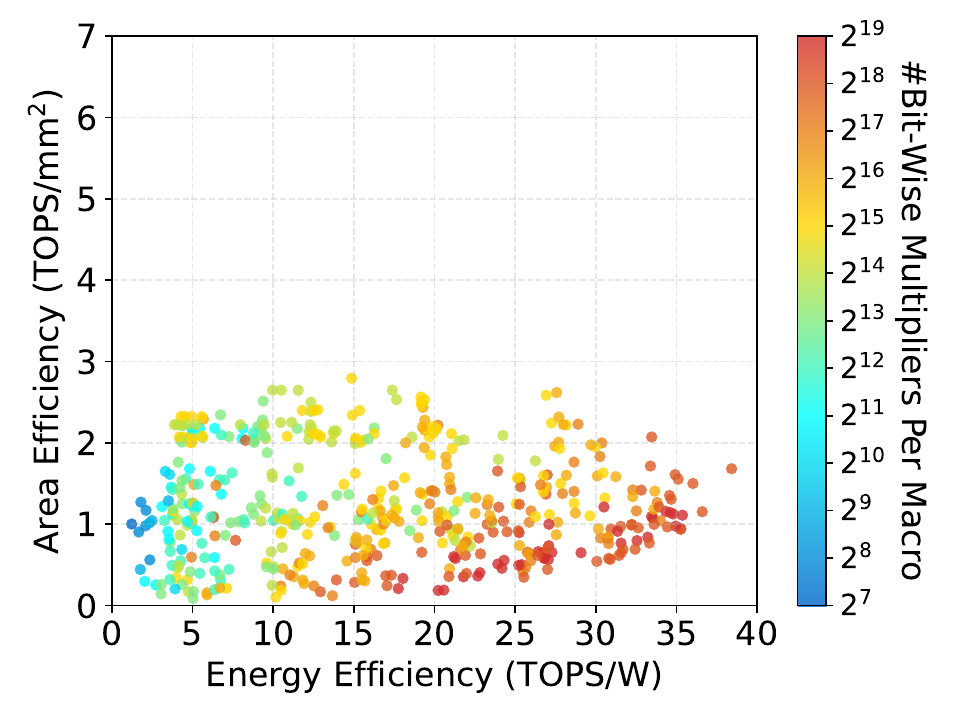}
        \caption{WS-Broadcast}
    \end{subfigure}
    \begin{subfigure}{0.49\linewidth}
        \centering
        \includegraphics[width=\linewidth]{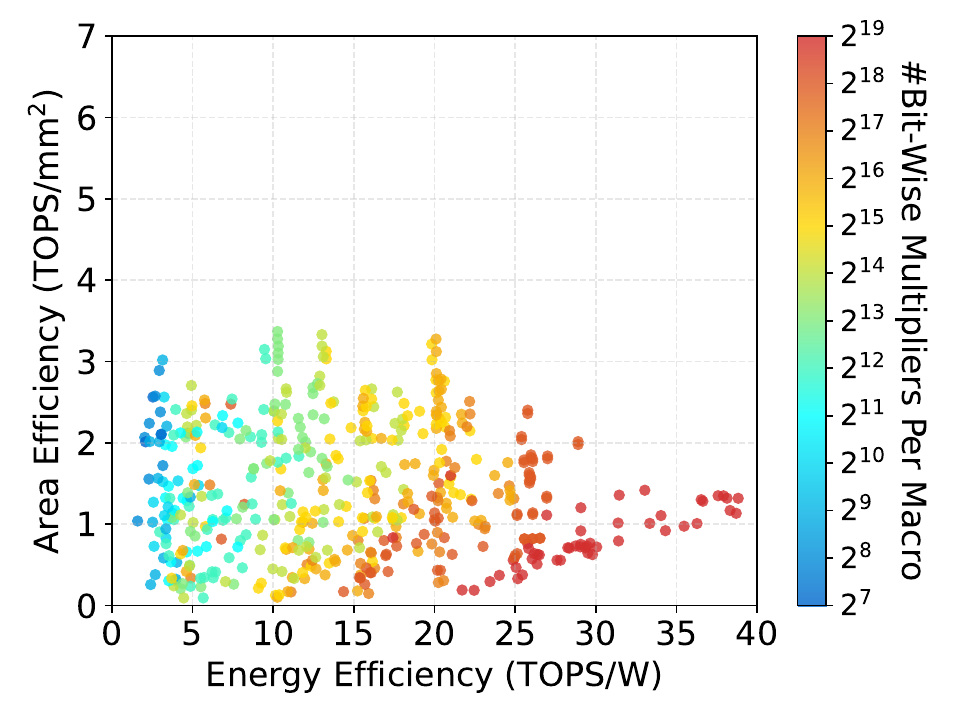}
        \caption{OS-Broadcast}
    \end{subfigure}
    
    \caption{The energy and area efficiency of candidate designs for CIM macro array with 512K bit-wise multipliers.}
    \label{fig:macro_selection}
\end{figure}

\begin{figure}[t]
    \centering
    \captionsetup{skip=2pt}
    \begin{subfigure}{0.49\linewidth}
        \centering
        \includegraphics[width=\linewidth]{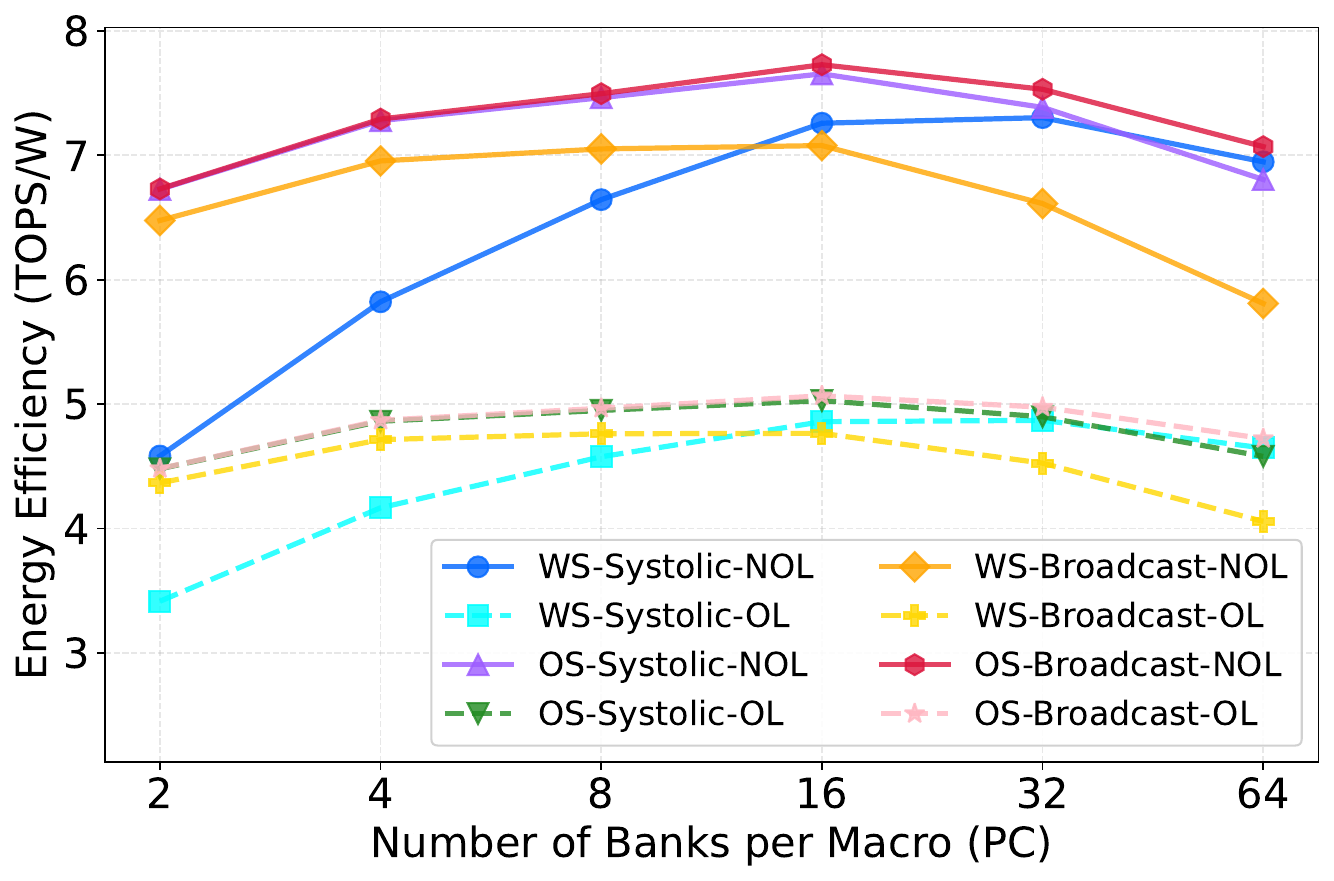}
        \caption{Energy efficiency}
    \end{subfigure}
    \begin{subfigure}{0.49\linewidth}
        \centering
        \includegraphics[width=\linewidth]{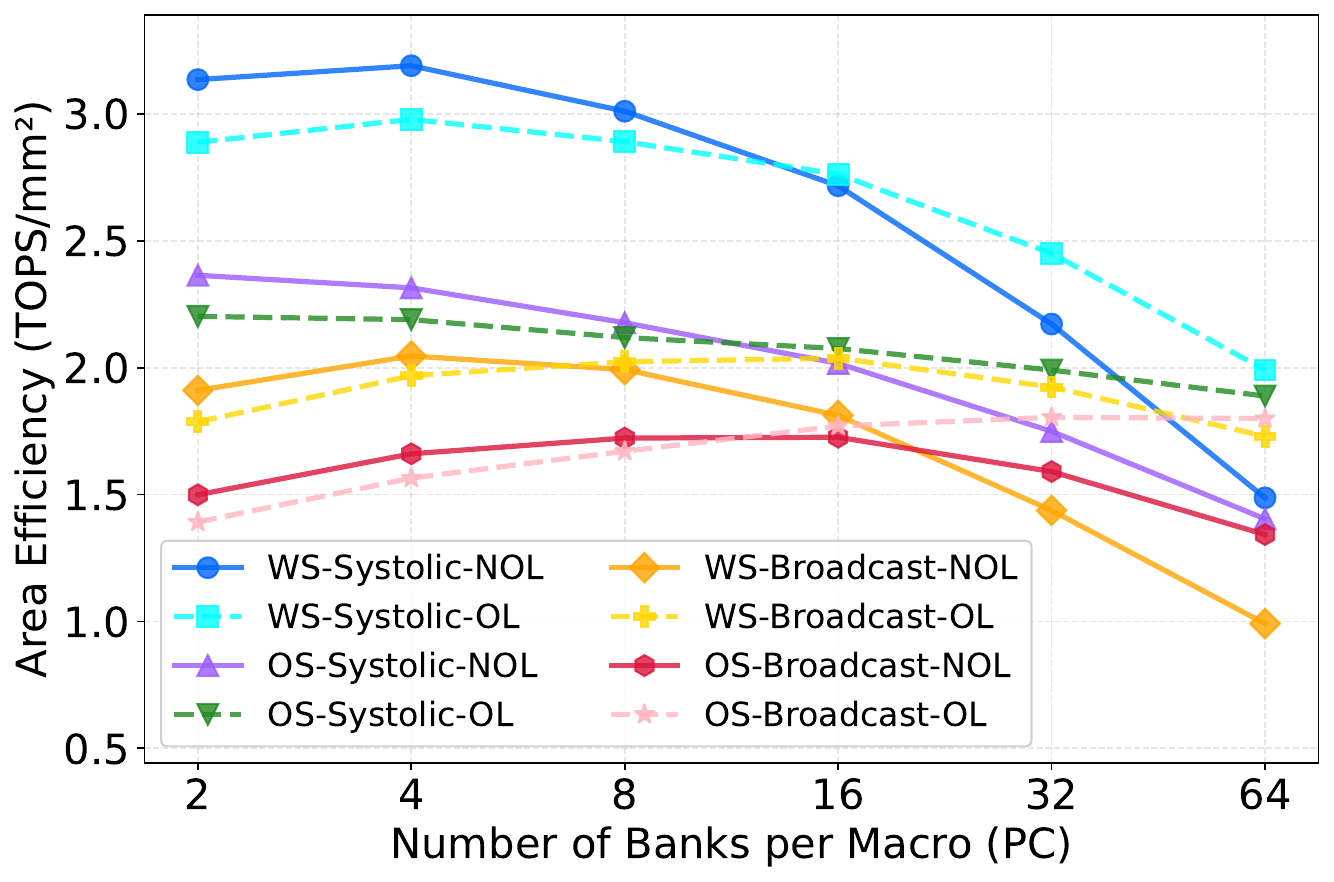}
        \caption{Area efficiency}
    \end{subfigure}
    
    \caption{Impact of CIM macro supporting Compute-I/O overlap on energy and area efficiency.}
    \label{fig:OL_efficiency}
    \vspace{-10pt}
\end{figure}

\noindent \textbf{Macro Selection.}
To determine the optimal macro compute capacity at the system level, we implement macro arrays with 512K bit-wise multipliers using different CIM macros, and evaluate the resulting design quality. This compute capacity also matches the maximum supported by the adopted CIM macro compiler. As shown in Figure~\ref{fig:macro_selection}, across four different dataflows, larger CIM macros provide better overall energy efficiency, since CIM macros dominate total power relative to peripheral logic. However, they deliver sub-optimal area efficiency due to significant internal broadcast overhead. Instead, medium-sized CIM macros achieve the highest area efficiency by balancing intra-macro implementation cost with inter-macro integration overhead.

\takeaway{Medium-sized CIM macros provide the best system-level trade-off between energy and area efficiency.}

\noindent \textbf{Compute-I/O Overlap.}
We investigate the impact of supporting simultaneous computation and weight update on accelerator area and energy efficiency. We adopt design macros that differ from the baseline 4-TOPS macro only in the number of banks ($PC$), and build 2$\times$4 macro arrays with different dataflows. As shown in Figure~\ref{fig:OL_efficiency}(a), enabling compute-I/O overlap generally leads to a ~25-35\% reduction in energy efficiency, primarily attributed to additional intra-macro circuitry. However, Figure~\ref{fig:OL_efficiency}(b) shows that compute-I/O overlap is beneficial for area efficiency when $PC$ is relatively large. This is because the weight update overhead is more pronounced when multiple banks in a CIM macro contend for the same I/O port.


\takeaway{Enabling compute-I/O overlap can improve area efficiency for bandwidth-constrained designs but comes at the cost of energy efficiency.}

\subsection{Case Study: LLM Inference}

We further evaluate \papername{} on a range of LLM workloads, including Qwen3-0.6B\cite{2505.09388}, LLaMA-3-8B/70B\cite{dubey2024llama}, and GPT-3 175B\cite{brown2020language}, targeting both edge and cloud deployment scenarios. The CIM core array is explored with an aggregate compute capacity upper bound of 20 TOPS. We measure the PPA trade-off as $latency^2 \cdot power\cdot area$ and present the optimal design points in Table~\ref{tab:case_study}. The latency is measured in terms of time-to-first-token at the prefill stage. 

As shown in Table~\ref{tab:case_study}, edge scenarios favor OS–Systolic–OL combined with moderate macro sizes and multi-core scaling. The overlap-enabled pipeline and column-wise weight sharing deliver considerable speedups while maintaining reasonable per-core area and power. Whereas for GPT-3 175B, WS–Systolic–NOL becomes preferable again due to its lower per-core area and power, which are crucial when scaling to tens of cores for long-sequence processing.
Notably, the optimal designs exhibit a low storage-compute ratio (LSL), which means memory capacity is less important for compute-intensive applications such as LLMs.
Overall, these results highlight the strong dependence of optimal design points on both model structure and deployment constraints, demonstrating the necessity of macro–array co-design and accurate PPA evaluation for achieving truly optimized CIM-based accelerators.

\section{Conclusion}
\label{sec:conclusion}

This paper introduces \papername{}, a systematic dataflow exploration framework for SRAM CIM accelerators. 
We identify oversimplified design choices and inaccurate modeling in prior CIM accelerator DSE frameworks. Motivated by these limitations, \papername{} constructs a comprehensive dataflow design space and rigorously evaluates the post-layout quality of candidate designs. Our results show that systolic interconnects are generally more efficient. In addition, whether to enable compute–I/O overlap and how to select the optimal macro capacity should be carefully tuned to the target application to balance energy efficiency and area.


\clearpage

\bibliographystyle{ACM-Reference-Format}
\bibliography{main}

@String{BIT = "{BIT}" }

@String{Computer = "{IEEE} Computer" }

@INPROCEEDINGS{DCIM-MediaTek,
  author={Shih, Ming-En and Hsieh, Shih-Wei and Tsai, Ping-Yuan and Lin, Ming-Hung and Tsung, Pei-Kuei and Chang, En-Jui and Liang, Jenwei and Chang, Shu-Hsin and Huang, Chung-Lun and Nian, You-Yu and Wan, Zhe and Kumar, Sushil and Xue, Cheng-Xin and Jedhe, Gajanan and Fujiwara, Hidehiro and Mori, Haruki and Chen, Chih-Wei and Huang, Po-Hua and Juan, Chih-Feng and Chen, Chung-Yi and Lin, Tsung-Yao and Wang, Ch and Chen, Chih-Cheng and Jou, Kevin},
  booktitle={2024 IEEE International Solid-State Circuits Conference (ISSCC)}, 
  title={20.1 NVE: A 3nm 23.2TOPS/W 12b-Digital-CIM-Based Neural Engine for High-Resolution Visual-Quality Enhancement on Smart Devices}, 
  year={2024},
  volume={67},
  number={},
  pages={360-362},
  doi={10.1109/ISSCC49657.2024.10454482}}

@online{dmatrix,
  author    = {d\text{-}Matrix},
  title     = {Corsair},
  url       = {https://www.d-matrix.ai/product/},
  urldate   = {2024/11},
  year      = {2024}
}

@online{houmo,
  author    = {Houmo},
  title     = {HaloDrive\textsuperscript{TM}30},
  url       = {https://www.houmoai.com/en/55/ProductType.html},
  urldate   = {2024/11},
  year      = {2024}
}

@inproceedings{chen2023autodcim,
  title={Autodcim: An automated digital cim compiler},
  author={Chen, Jia and Tu, Fengbin and Shao, Kunming and Tian, Fengshi and Huo, Xiao and Tsui, Chi-Ying and Cheng, Kwang-Ting},
  booktitle={2023 60th ACM/IEEE Design Automation Conference (DAC)},
  pages={1--6},
  year={2023},
  organization={IEEE}
}

@inproceedings{zhang2024arctic,
  title={Arctic: Agile and robust compute-in-memory compiler with parameterized int/fp precision and built-in self test},
  author={Zhang, Hongyi and Zhu, Haozhe and He, Siqi and Li, Mengjie and Wang, Chengchen and Xiong, Xiankui and Tian, Haidong and Zeng, Xiaoyang and Chen, Chixiao},
  booktitle={2024 Design, Automation \& Test in Europe Conference \& Exhibition (DATE)},
  pages={1--6},
  year={2024},
  organization={IEEE}
}

@inproceedings{shao2025syndcim,
  title={Syndcim: A performance-aware digital computing-in-memory compiler with multi-spec-oriented subcircuit synthesis},
  author={Shao, Kunming and Tian, Fengshi and Wang, Xiaomeng and Zheng, Jiakun and Chen, Jia and He, Jingyu and Wu, Hui and Chen, Jinbo and Guan, Xihao and Deng, Yi and others},
  booktitle={2025 Design, Automation \& Test in Europe Conference (DATE)},
  pages={1--7},
  year={2025},
  organization={IEEE}
}

@inproceedings{wang2025damil,
  title={DAMIL-DCIM: A Digital CIM Layout Synthesis Framework with Dataflow-Aware Floorplan and MILP-Based Detailed Placement},
  author={Wang, Chuyu and Hu, Ke and Yang, Fan and Zhu, Keren and Zeng, Xuan},
  booktitle={2025 Design, Automation \& Test in Europe Conference (DATE)},
  pages={1--7},
  year={2025},
  organization={IEEE}
}

@inproceedings{diao2025sega,
  title={SEGA-DCIM: Design Space Exploration-Guided Automatic Digital CIM Compiler with Multiple Precision Support},
  author={Diao, Haikang and Zhang, Haoyi and Song, Jiahao and Luo, Haoyang and Lin, Yibo and Wang, Runsheng and Wang, Yuan and Tang, Xiyuan},
  booktitle={2025 Design, Automation \& Test in Europe Conference (DATE)},
  pages={1--7},
  year={2025},
  organization={IEEE}
}

@article{zhu2023mnsim,
  title={Mnsim 2.0: A behavior-level modeling tool for processing-in-memory architectures},
  author={Zhu, Zhenhua and Sun, Hanbo and Xie, Tongxin and Zhu, Yu and Dai, Guohao and Xia, Lixue and Niu, Dimin and Chen, Xiaoming and Hu, Xiaobo Sharon and Cao, Yu and others},
  journal={IEEE Transactions on Computer-Aided Design of Integrated Circuits and Systems},
  volume={42},
  number={11},
  pages={4112--4125},
  year={2023},
  publisher={IEEE}
}

@article{lee2024neurosim,
  title={Neurosim v1. 4: Extending technology support for digital compute-in-memory toward 1nm node},
  author={Lee, Junmo and Lu, Anni and Li, Wantong and Yu, Shimeng},
  journal={IEEE Transactions on Circuits and Systems I: Regular Papers},
  volume={71},
  number={4},
  pages={1733--1744},
  year={2024},
  publisher={IEEE}
}

@inproceedings{andrulis2024cimloop,
  title={CiMLoop: A flexible, accurate, and fast compute-in-memory modeling tool},
  author={Andrulis, Tanner and Emer, Joel S and Sze, Vivienne},
  booktitle={2024 IEEE International Symposium on Performance Analysis of Systems and Software (ISPASS)},
  pages={10--23},
  year={2024},
  organization={IEEE}
}

@article{wu2025exploiting,
  title={Exploiting the Memory-Compute-Coupling Feature for CIM Accelerator Design Optimization},
  author={Wu, Yongkun and Wang, Xiaomeng and Chen, Jia and Zhu, Zhenhua and He, Jingyu and Dong, Pingcheng and Tan, Yonghao and Zhao, Xin and Chang, Liang and Wang, Yu and others},
  journal={IEEE Transactions on Computer-Aided Design of Integrated Circuits and Systems},
  year={2025},
  publisher={IEEE}
}

@article{qi2025cimflow,
  title={CIMFlow: An Integrated Framework for Systematic Design and Evaluation of Digital CIM Architectures},
  author={Qi, Yingjie and Yang, Jianlei and Wang, Yiou and Wang, Yikun and Wang, Dayu and Tang, Ling and Duan, Cenlin and He, Xiaolin and Zhao, Weisheng},
  journal={arXiv preprint arXiv:2505.01107},
  year={2025}
}

@inproceedings{sun2023analog,
  title={Analog or digital in-memory computing? benchmarking through quantitative modeling},
  author={Sun, Jiacong and Houshmand, Pouya and Verhelst, Marian},
  booktitle={2023 IEEE/ACM International Conference on Computer Aided Design (ICCAD)},
  pages={1--9},
  year={2023},
  organization={IEEE}
}

@inproceedings{zhu2025leveraging,
  title={Leveraging compute-in-memory for efficient generative model inference in tpus},
  author={Zhu, Zhantong and Li, Hongou and Ren, Wenjie and Wu, Meng and Ye, Le and Huang, Ru and Jia, Tianyu},
  booktitle={2025 Design, Automation \& Test in Europe Conference (DATE)},
  pages={1--7},
  year={2025},
  organization={IEEE}
}

@inproceedings{chen2023systolic,
  title={A systolic computing-in-memory array based accelerator with predictive early activation for spatiotemporal convolutions},
  author={Chen, Xiaofeng and Guo, Ruiqi and Yue, Zhiheng and Hu, Yang and Liu, Leibo and Wei, Shaojun and Yin, Shouyi},
  booktitle={2023 IEEE 5th International Conference on Artificial Intelligence Circuits and Systems (AICAS)},
  pages={1--5},
  year={2023},
  organization={IEEE}
}

@article{zhan2024gslp,
  title={GSLP-CIM: A 28-nm globally systolic and locally parallel CNN/transformer accelerator with scalable and reconfigurable eDRAM compute-in-memory macro for flexible dataflow},
  author={Zhan, Yi and Yu, Wei-Han and Un, Ka-Fai and Martins, Rui P and Mak, Pui-In},
  journal={IEEE Transactions on Circuits and Systems I: Regular Papers},
  year={2024},
  publisher={IEEE}
}

@article{dubey2024llama,
  title={The llama 3 herd of models},
  author={Dubey, Abhimanyu and Jauhri, Abhinav and Pandey, Abhinav and Kadian, Abhishek and Al-Dahle, Ahmad and Letman, Aiesha and Mathur, Akhil and Schelten, Alan and Yang, Amy and Fan, Angela and others},
  journal={arXiv e-prints},
  pages={arXiv--2407},
  year={2024}
}

@software{cimcompiler,
  author = {Hui, Richard},
  title = {{CIAL-CIMCompiler}},
  url = {https://github.com/Richard-Hui/CIAL-CIMCompiler},
  year = {2025}
}

@inproceedings{fujiwara20225,
  title={A 5-nm 254-TOPS/W 221-TOPS/mm 2 fully-digital computing-in-memory macro supporting wide-range dynamic-voltage-frequency scaling and simultaneous MAC and write operations},
  author={Fujiwara, Hidehiro and Mori, Haruki and Zhao, Wei-Chang and Chuang, Mei-Chen and Naous, Rawan and Chuang, Chao-Kai and Hashizume, Takeshi and Sun, Dar and Lee, Chia-Fu and Akarvardar, Kerem and others},
  booktitle={2022 IEEE International Solid-State Circuits Conference (ISSCC)},
  volume={65},
  pages={1--3},
  year={2022},
  organization={IEEE}
}

@inproceedings{fujiwara202434,
  title={34.4 A 3nm, 32.5 TOPS/W, 55.0 TOPS/mm 2 and 3.78 Mb/mm 2 fully-digital compute-in-memory macro supporting INT12$\times$ INT12 with a parallel-MAC architecture and foundry 6T-SRAM bit cell},
  author={Fujiwara, Hidehiro and Mori, Haruki and Zhao, Wei-Chang and Khare, Kinshuk and Lee, Cheng-En and Peng, Xiaochen and Joshi, Vineet and Chuang, Chao-Kai and Hsu, Shu-Huan and Hashizume, Takeshi and others},
  booktitle={2024 IEEE International Solid-State Circuits Conference (ISSCC)},
  volume={67},
  pages={572--574},
  year={2024},
  organization={IEEE}
}

@article{brown2020language,
  title={Language models are few-shot learners},
  author={Brown, Tom and Mann, Benjamin and Ryder, Nick and Subbiah, Melanie and Kaplan, Jared D and Dhariwal, Prafulla and Neelakantan, Arvind and Shyam, Pranav and Sastry, Girish and Askell, Amanda and others},
  journal={Advances in neural information processing systems},
  volume={33},
  pages={1877--1901},
  year={2020}
}

@misc{2505.09388,
Author = {An Yang and Anfeng Li and Baosong Yang and Beichen Zhang and Binyuan Hui and Bo Zheng and Bowen Yu and Chang Gao and Chengen Huang and Chenxu Lv and Chujie Zheng and Dayiheng Liu and Fan Zhou and Fei Huang and Feng Hu and Hao Ge and Haoran Wei and Huan Lin and Jialong Tang and Jian Yang and Jianhong Tu and Jianwei Zhang and Jianxin Yang and Jiaxi Yang and Jing Zhou and Jingren Zhou and Junyang Lin and Kai Dang and Keqin Bao and Kexin Yang and Le Yu and Lianghao Deng and Mei Li and Mingfeng Xue and Mingze Li and Pei Zhang and Peng Wang and Qin Zhu and Rui Men and Ruize Gao and Shixuan Liu and Shuang Luo and Tianhao Li and Tianyi Tang and Wenbiao Yin and Xingzhang Ren and Xinyu Wang and Xinyu Zhang and Xuancheng Ren and Yang Fan and Yang Su and Yichang Zhang and Yinger Zhang and Yu Wan and Yuqiong Liu and Zekun Wang and Zeyu Cui and Zhenru Zhang and Zhipeng Zhou and Zihan Qiu},
Title = {Qwen3 Technical Report},
Year = {2025},
Eprint = {arXiv:2505.09388},
}

@article{guo202428,
  title={A 28-nm 64-kb 31.6-TFLOPS/W digital-domain floating-point-computing-unit and double-bit 6T-SRAM computing-in-memory macro for floating-point CNNs},
  author={Guo, An and Xi, Chen and Dong, Fangyuan and Pu, Xingyu and Li, Dongqi and Zhang, Jingmin and Dong, Xueshan and Gao, Hui and Zhang, Yiran and Wang, Bo and others},
  journal={IEEE Journal of Solid-State Circuits},
  volume={59},
  number={9},
  pages={3032--3044},
  year={2024},
  publisher={IEEE}
}

@inproceedings{li2021openbox,
  title={Openbox: A generalized black-box optimization service},
  author={Li, Yang and Shen, Yu and Zhang, Wentao and Chen, Yuanwei and Jiang, Huaijun and Liu, Mingchao and Jiang, Jiawei and Gao, Jinyang and Wu, Wentao and Yang, Zhi and others},
  booktitle={Proceedings of the 27th ACM SIGKDD conference on knowledge discovery \& data mining},
  pages={3209--3219},
  year={2021}
}
\end{document}